\documentclass[10pt,b5paper,twoside,onecolumn]{scrartcl}
\usepackage[cp1250]{inputenc}
\usepackage{color,fancyvrb}
\usepackage{graphicx}
\usepackage{hyperref}
\usepackage{fancyhdr}
\usepackage{setspace}
\usepackage{float}
\usepackage{pdfpages}
\usepackage{verbatim}
\usepackage{lmodern} 
\usepackage{amsthm}
\usepackage{amssymb}
\usepackage{amsmath}
\usepackage{multicol}
\usepackage{mathpazo}
\usepackage[T1]{fontenc}
\usepackage[titles]{tocloft}
\begin{document}
	
\pagestyle{fancyplain}
\fancyhf{}
\fancyhead[LE]{\textit{A Memetic Algorithm for the Minimum Conductance Graph Partitioning Problem}}
\fancyhead[RO]{\textit{D. Chalupa}}
\fancyfoot[C]{\thepage}
\fancypagestyle{plain}
{
	\fancyhf{} 
	\renewcommand{\headrulewidth}{0pt} 
	\renewcommand{\footrulewidth}{0pt}
}

\thispagestyle{empty}
	
\begin{center}\textbf{\LARGE\sffamily\noindent
A Memetic Algorithm for the Minimum Conductance Graph Partitioning Problem
}\end{center}

\begin{center}\textbf{\large\sffamily\noindent
[working paper]
}
\end{center}

\begin{center}{\large\sffamily\noindent David Chalupa}\end{center}
\begin{center}
{
\noindent
Computer Science\\ School of Engineering and Computer Science\\
University of Hull\\
Cottingham Road\\
Hull HU6 7RX, UK\\
Email: \texttt{D.Chalupa@hull.ac.uk}
}
\end{center}

\vspace{30pt}
	
\paragraph{Abstract.} The minimum conductance problem is an NP-hard graph partitioning problem.
Apart from the search for bottlenecks in complex networks, the problem is very closely related to the popular area of network community detection.
In this paper, we tackle the minimum conductance problem as a pseudo-Boolean optimisation problem and propose a memetic algorithm to solve it.
An efficient local search strategy is established.
Our memetic algorithm starts by using this local search strategy with different random strings to sample a set of diverse initial solutions.
This is followed by an evolutionary phase based on a steady-state framework and two intensification subroutines.
We compare the algorithm to a wide range of multi-start local search approaches and classical genetic algorithms with different crossover operators.
The experimental results are presented for a diverse set of real-world networks. 
These results indicate that the memetic algorithm outperforms the alternative stochastic approaches.


\section{Introduction}

The minimum conductance problem finds its applications in \textit{network community detection} \cite{leskovec}, as well as in general \textit{graph clustering} \cite{brandes2003experiments,graphclustering}. It is also related to other graph partioning problems such as clique covering \cite{ist14}. In the minimum conductance problem, the aim is to divide the vertex set into two subsets such that the ``relative connectivity'' of these two subsets is minimised. In some studies, it is also referred to as the \textit{sparsest cut problem} \cite{OptimizingStreamProcessingGraphPartitioning}.

Figure 1 illustrates the problem for a small $52$-vertex social network. The problem aims at partitioning the vertex set into two partitions, minimising the ratio of the number of edges connecting vertices in different partitions to the number of all edges incident to the vertices incident with vertices of one of the partitions.

The problem was shown to be NP-hard by \v{S}\'{i}ma and Schaeffer approximately a decade ago \cite{nphardclustering2}. Despite this fact, the problem has been overlooked as an optimisation problem. Leskovec et al used the conductance metric to evaluate the quality of network communities obtained by different types of algorithms \cite{communitylarge,leskovec}. In this context, conductance is used as a measure of how well a community is separated from the rest of the graph.

The minimum conductance problem can naturally be represen\-ted as a $0$-$1$ optimisation problem, in the form of a pseudo-Boolean function \cite{BasicPseudoBooleanProgramming}. Conductance is defined for any partitioning, apart from those, which contain a partition consisting solely of isolated vertices. In the following, we will show that conductance can be expressed as a non-linear pseudo-Boolean function.

This outlines a link between this problem and a wide range of prominent problems in non-linear pseudo-Boolean optimisation such as
maxSAT \cite{BestImprovingVersusFirstImproving}, maxkSAT \cite{ConstantTimeSteepestDescent},
NK-landscapes \cite{SecondPartialDerivativesNKLandscapes,NKLandscapesNPHard} or other tunable rugged objective functions \cite{TunablyRugged}.

For many of these problems, efficient partition crossovers \cite{PartitionCrossoverPseudoboolean}, neighbourhood search strategies \cite{ImprovingMovesPseudoboolean} and landscape analysis techniques have been applied \cite{CrossoverNetworks,ChallengingPseudoboolean}. These results generally rely on the $k$-bounded nature of the pseudo-Boolean functions, i.e. these functions can be expressed as a sum of terms such that each term depends on at most $k$ Boolean inputs \cite{GrayBoxMkLandscapes}. In this paper, we lay the foundations for a similar study of the minimum conductance problem.

\begin{figure}
\begin{center}
\includegraphics[scale=0.22]{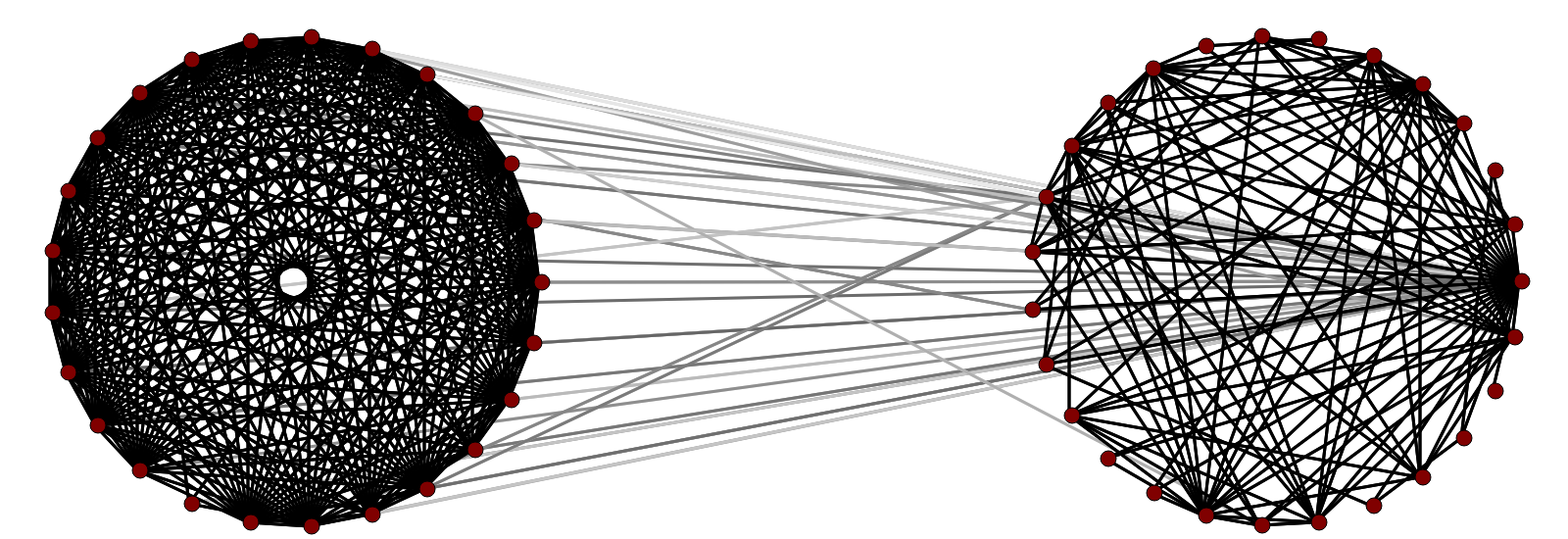}
\caption{An example of a low conductance graph partitioning found for a $52$-vertex social network sample $soc\_52$. The conductance value for this partitioning is $0.13108614$. The drawing shows that the solution partitions the network into two dense clusters, connected relatively sparsely.}
\end{center}
\end{figure}

We have outlined that the minimum conductance problem is closely related to community detection. Over the years, many different strategies of community detection have been identified \cite{ChineseWhispersClustering,blondel2008fast,EvolutionaryCommunityDiscoveryDynamicNetworks,CommunityDetectionInGraphs,ComplexNetworkClusteringParticleSwarm,MultilevelCommunityDetection,EnergyModelsGraphClustering,CommunitiesComplexNetworks}. Community structure of networks in many domains has been studied \cite{CombinatorialModelCommunityStructure}, including \textit{social and biological networks} \cite{communitystructure1}. Related but more general concepts include graph clustering \cite{stochasticlocalclustering,graphclustering,LocalClusteringMassiveGraphs} and graph mining \cite{graphmining1,graphmining3}. These concepts find their applications in various areas, including \textit{social media} \cite{CommunityDetectionInSocialMedia}, \textit{web communities} \cite{webcommunities}, \textit{cyberattack detection} \cite{DetectingCoalitionHitInflationAttacks}, as well as \textit{functional module detection} for protein-protein interaction networks \cite{EvolutionaryProteinProteinInteractions}.

\newpage
This problem is also closely related to the analysis of \textit{scale-free} \cite{barabasialbert2,dorogmendes} and \textit{small-world networks} \cite{smallworlds}, as well as the \textit{hierarchy} observed in real-world networks \cite{hierarchylocal,hierarchynetworks}.

\textit{Contributions.} In this paper, we propose a \textit{steady-state adaptive memetic algorithm} (StS AMA) for the minimum conductance problem. We also explore the potential of classical genetic algorithms and several local search strategies in solving the problem. We also derive an \textit{efficient neighbourhood exploration strategy}, along with an efficient way of conductance recalculation in local search. Our experimental results indicate that StS AMA outperforms genetic algorithms with both one-point and uniform crossovers, as well as local search and randomised local search strategies. The experiments are provided using six different search algorithms applied to both \textit{social networks} and \textit{protein-protein interaction networks}.

In Section 2, we describe the minimum conductance problem and conductance as a pseudo-Boolean function. In Section 3, we describe our efficient neighbourhood exploration strategy and provide an overview of local search and crossover-based evolutionary algorithms we have used to tackle the problem. In Section 4, we describe our StS AMA. In Section 5, we present the obtained experimental results. Last but not least, in Section 6, we conclude the work and provide a discussion on its outcome and related open problems.

\section{The Minimum Conductance Problem}

In this section, we first describe the minimum conductance problem as a graph problem. Secondly, we formulate conductance as a pseudo-Boolean function and discuss the ways for its efficient calculation and recalculation in local search and evolutionary algorithms.

\subsection{Problem Definition}

Let $G = [V,E]$ be a connected undirected graph and let $S \subset V$ be a subset of its vertices such that $S \neq \emptyset$ and $S \neq V$.
Then, the \textit{volume} of this subset $S$ is defined as follows:

\begin{equation}
Vol(S) = \sum_{v \in S} deg(v),
\end{equation}

\noindent
where $deg(v)$ is the degree of $v \in V$, i.e. the number of its neighbours.

Then, the \textit{conductance} $\Phi(S)$ of a partitioning of $V$ into the sets $S$ and $V \backslash S$ is defined using the following formula:

\begin{equation}
\Phi(S) = \frac{c_G(S)}{\min\{Vol(S), Vol(V \backslash S)\}}.
\end{equation}

\noindent
where

\begin{equation}
c_G(S) = \sum_{v \in S} deg_{V \backslash S}(v) = \sum_{v \in V \backslash S} deg_{S}(v),
\end{equation}

\noindent
with $deg_{V \backslash S}(v) = |\{ \{v,v'\}: ~ v' \in V \backslash S \}|$ being the number of neighbours of $v$ in the set $V \backslash S$.

One can see that $\Phi(S)$ is not defined if $Vol(S) = 0$ or $Vol(V \backslash S) = 0$. For a connected graph, that holds if and only if $S = \emptyset$ or $S = V$.

For any other solutions, we have that $Vol(S) > 0$ and $Vol(V \backslash S) > 0$. This implies that $\Phi(S)$ can be rewritten as:

\begin{equation}
\Phi(S) = \max\left\{\frac{c_G(S)}{Vol(S)},\frac{c_G(S)}{Vol(V \backslash S)}\right\}.
\end{equation}

\subsection{Conductance as a Pseudo-Boolean Function and its Properties}

Variable $c_G(S)$ represents the sum of the numbers of neighbours of vertices $v \in S$, which are in $V \backslash S$ (or vice versa). Hence, $\Phi(S)$ can be further transformed into the following form:

\begin{equation}
\Phi(S) = \max\left\{\frac{\sum_{v \in S} deg_{V \backslash S}(v)}{\sum_{v \in S} deg(v)}, \frac{\sum_{v \in V \backslash S} deg_{S}(v)}{\sum_{v \in V \backslash S} deg(v)}\right\}.
\end{equation}

\noindent
This shows that the conductance represents a maximum of two pseudo-Bool\-ean functions. In each of these functions, the numerator is a sum of values, each of which depends on $deg(v)$ variables, depending on the degree of $v$. However, the denominators influence all of the elements of the sum. With each move of a vertex from $S$ to $V \backslash S$, or vice versa, the volumes of the sets are changed.

In other words, with each bit flip in a $0$-$1$ representation of the problem, the values of all elements of the sums can be potentially be changed.

Let $S' = S \cup \{v\}$ or $S' = S \backslash \{v\}$ and let $c = c_G(S') - c_G(S)$. Then, an improvement or a stagnation (i.e. $\Phi(S') \leq \Phi(S)$) will be obtained if and only if:

\begin{equation}
\frac{\min\{Vol(S), Vol(V \backslash S)\}}{\min\{Vol(S'), Vol(V \backslash S')\}} \Phi(S) - \Phi(S) + \frac{c}{\min\{Vol(S'), Vol(V \backslash S')\}} \leq 0,
\end{equation}

\noindent
which can further be transformed to:

\begin{equation}
\min\{Vol(S), Vol(V \backslash S)\}  - \min\{Vol(S'), Vol(V \backslash S')\} + \frac{c}{\Phi(S)} \leq 0.
\end{equation}

\noindent
One can see that this condition can be fulfilled in multiple ways and does not seem to be as straightforward as for the case of $k$-bounded pseudo-Boolean optimisation \cite{GrayBoxMkLandscapes}. In fact, the numerator of conductance is $k$-bounded, where $k$ is equivalent to the maximum degree of our graph. However, the value in the denominator can be potentially changed in all $|V|$ components.

Remarkably, for $k$-bounded functions, improving moves can be identified in $\mathcal{O}(1)$ time without actually scanning the neighbourhood \cite{ImprovingMovesPseudoboolean}. It is worth noting that for the minimum conductance problem, it is possible to find moves, for which

\begin{equation}
\frac{c}{\min\{Vol(S'), Vol(V \backslash S')\}} < 0,
\end{equation}

\noindent
in $\mathcal{O}(1)$ time. However, the volume ratio $\frac{\min\{Vol(S), Vol(V \backslash S)\}}{\min\{Vol(S'), Vol(V \backslash S')\}}$ seems to complicate the situation for the minimum conductance problem.

Therefore, it remains open whether improving moves can be found in $\mathcal{O}(1)$ time for minimum conductance problem. However, we will show in the next section that it is possible to recalculate the conductance for a single move in $\mathcal{O}(1)$ time. This leads to efficient randomised local search strategies, as well as systematic local search, which can be used to scan the neighbourhood fully in $\mathcal{O}(n)$ time.

\section{Local Search Strategies and Genetic Algorithms}

In this section, we first present our neighbourhood exploration strategy. Next, we describe the three local search and two genetic algorithms that we used in our experimental evaluations. The next section will then follow up with the description of StS AMA.

\subsection{Neighbourhood Exploration Strategy}

The conductance for a partitioning $S$ can be computed in $\mathcal{O}(m)$ time for a graph on $m$ edges, by iterating over the edges of the graph. At the same time, the current value of $c_G(S)$, and degrees $deg_S(v)$ and $deg_{V \backslash S}(v)$ can be calculated and stored as auxiliary data.

Let $S' = S \cup \{v\}$. Then, the following formulas can be used to recalculate the conductance in $\mathcal{O}(1)$ time:

\begin{equation}
c_G(S') = c_G(S) - deg_S(v) + deg_{V \backslash S}(v),
\end{equation}

\begin{equation}
Vol(S') = Vol(S) - deg(v),
\end{equation}

\begin{equation}
Vol(V \backslash S') = Vol(V \backslash S) + deg(v).
\end{equation}

\noindent
If the move is accepted, then for each neighbour $w$ of our vertex $v$, the auxiliary degrees can be updated as follows:

\begin{equation}
deg_{S'}(w) = deg_{S}(w) - 1,
\end{equation}

\begin{equation}
deg_{V \backslash S'}(w) = deg_{V \backslash S}(w) + 1.
\end{equation}

\subsection{Local Search and Genetic Algorithms}

Many local search approaches may be used to solve this type of a problem \cite{StochasticLocalSearch}, even though it seems that none of them have been applied yet. In the following, we describe the algorithms we have used to tackle the minimum conductance problem.

\begin{table}
\begin{center}
Algorithm 1: An Adaptive Local Search Algorithm (ALS$^1$) for the Minimum Conductance Graph Partitioning Problem\vspace{5pt}\\
\begin{tabular}{l|l}
& Output: best configuration $S_{best}$ found\\\hline
1  & $p_s = 1/2$, $\Phi_{best} = \infty$\\
2  & while stopping criteria are not met\\
3  & \hspace{10pt}set each bit of $S$ to $1$ with probability $p_s$\\
4  & \hspace{10pt}improve $S$ using LS$^1$ until the local optimum is reached\\
5  & \hspace{10pt}if $\Phi(S) \leq \Phi_{best}$\\
6  & \hspace{10pt}\hspace{10pt}$\Phi_{best} = \Phi(S)$\\
7  & \hspace{10pt}\hspace{10pt}$p_s = p_s / 2$\\
8  & \hspace{10pt}else\\
9  & \hspace{10pt}\hspace{10pt}$p_s = 1/2$\\
10 & return $S_{best}$\\
\end{tabular}
\end{center}
\end{table}

\textit{Local search algorithm LS$^1$.} This is a simple steepest descent search algorithm, which uses the strategy described above to test each of the possible bit flips in $\mathcal{O}(1)$ time per vertex. The algorithm starts with a random bit string and chooses the best bit flip in each iteration. It stops whenever the best solution in the neighbourhood is not better than the current solution. Therefore, LS$^1$ guarantees that a local optimum is reached. If a local optimum is found, the local search is restarted.

\textit{Adaptive local search algorithm ALS$^1$.} This is an extension of LS$^1$. The pseudocode of ALS$^1$ is given in Algorithm 1. The adaptive component relies on a gradual lowering of the probability of a $1$-bit being generated in the initial solution. This way, we ensure that the algorithm also searches for asymmetric partitionings, which may be hard to reach if the initial solution contains $0$-bits and $1$-bits with equal probabilities. This is accomplished by starting with generating the initial solution by assigning $1$-bits with probability $p_s = 1/2$. Then, in the next restart of the local search, $p_s$ is halved to $1/4$. If the best solution sampled by the last run of the local search has not improved the best solution found so far, $p_s$ is reset to $1/2$.

\textit{Adaptive randomised local search algorithm ARLS$^{1,2}$.} This algorithm is a randomised local search approach, which allows both moves of a single vertex between partitions, as well as exchanges of vertices between them. In each iteration, a move of one vertex or a move of two different vertices is tested using the approach described above. The move is accepted if the new conductance is at least as good as the current conductance. Both the test of a move and an update after the move take $\mathcal{O}(\Delta)$ time, where $\Delta$ is the maximum degree of a vertex.

\textit{Adaptive genetic algorithm with one-point crossover AGA-1PX.} AGA-1PX is a relatively standard variant of a genetic algorithm with a one-point crossover. It uses a population $P$ of $p$ individuals initially generated at random. A $1$-bit is placed in the initial solution with probability $p_s$ and a $0$-bit is placed into it with probability $1 - p_s$. The initial value $p_s$ is set to $1/2$. Similarly to ALS$^1$, $p_s$ is iteratively halved until no improvement in the initial conductance is obtained. This occurs whenever $10^6$ candidate solutions have been generated without improvement of the best solution found so far. AGA-1PX then restarts the search with halved $p_s$. In the evolutionary phase, tournament of size $t$ is used to select two parents. Two offspring are then generated using the one-point crossover. Next, mutation is performed on each offspring. Each bit in the solution is flipped with probability $1 / n$, where $n$ is the number of vertices in the graph. This is repeated until $p$ new solutions are generated. We use a slightly elitist replacement strategy. All of the individuals are replaced with the offspring, apart from the best individual in the population.

\textit{Adaptive genetic algorithm with uniform crossover AGA-UX.} This algorithm has almost exactly the same structure as AGA-1PX. The only difference is that for each pair of parents, one offspring is generated by the uniform crossover. In the uniform crossover, each bit is taken from the first parent with probability $1/2$ and from the second parent otherwise \cite{uniformcrossover}. The selection, mutation and replacement strategies are exactly the same as in AGA-1PX.

\section{Steady-state Adaptive Memetic Algorithm}

In this section, we introduce the steady-state adaptive memetic algorithm (StS AMA) that we propose for the problem.

Memetic algorithms have been widely used and their efficiency is related to properties of the fitness landscape \cite{chen2011multi,competentmemetic,MemeticAlgorithmsAndLandscapes,memeticpopulationmanagement}. Memetic algorithms have also been previously used to solve different graph partitioning problems \cite{MemeticGraphPartitioning}.

The pseudocode of our StS AMA is given in Algorithm 2. The algorithm uses a population of individuals, which represent local optima. In the beginning, the initial population is generated adaptively at random, and is improved by the application of the systematic local search algorithm LS$^1$. The adaptive component addresses the observation that low-conductance partitionings of real-world networks often seem to have an unbalanced structure. Therefore, each initial individual $P_i$ is generated uniformly at random, with each bit set to $1$ with probability $p_s = 1/2$ and to $0$ otherwise. As the next step, $p_s$ is lowered to $1/4$ to allow the exploration of unbalanced solutions. This process is described in steps 2-8 and stops whenever another lowering of $p_s$ leads to a worse solution than the previous solution found.

This is followed by an evolutionary procedure. In each generation, two parents $P_1$ and $P_2$ are chosen by a tournament of size $t$ in step 11. In step 12, their offspring $O_1$ and $O_2$ are created using one-point crossover. In steps 13-14, RLS$^{1,2}$ and LS$^1$ are consecutively applied to improve $O_1$ and $O_2$. Note that after step 14, both $O_1$ and $O_2$ represent locally optimal partitionings. In steps 15-16, $O_1$ and $O_2$ are used to replace the worst individuals in the population. This process is iterated until a stopping criterion is met. In our experiments, we will simply use a time limit as this criterion.

\begin{table}
\begin{center}
Algorithm 2: A Steady-state Adaptive Memetic Algorithm (StS AMA) for the Minimum Conductance Graph Partitioning Problem\vspace{5pt}\\
\begin{tabular}{l|l}
& Input: population size $p$, tournament size $t$,\\
& local search length $l$\\
& Output: best configuration $P_{best}$ found\\\hline
1  & for $i = 1...p$\\
2  & \hspace{10pt}$p_s = 1/2$\\
3  & \hspace{10pt}do\\
4  & \hspace{10pt}\hspace{10pt}set each bit of a candidate for individual $P_i$\\
    & \hspace{10pt}\hspace{10pt}to $1$ with probability $p_s$\\
5  & \hspace{10pt}\hspace{10pt}improve the candidate for individual $P_i$ using LS$^1$\\
    & \hspace{10pt}\hspace{10pt}until the local optimum is reached\\
6  & \hspace{10pt}\hspace{10pt}$p_s = p_s / 2$\\
7  & \hspace{10pt}while the current candidate for $P_i$ is at least as good \\
    & \hspace{10pt}as the best of the previous candidates\\
8  & \hspace{10pt}set the best candidate sampled in steps 3-7 as the\\
   & \hspace{10pt}individual $P_i$\\
9  & $P = \{P_1, P_2, ..., P_p\}$\\
10  & while stopping criteria are not met\\
11 & \hspace{10pt}pick parents $P_{p_1}$ and $P_{p_2}$ such that $p_1 \neq p_2$ using\\
    & \hspace{10pt}a tournament of size $t$\\
12  & \hspace{10pt}create the offspring $O_1$ and $O_2$ using\\
    & \hspace{10pt}one-point crossover\\
13  & \hspace{10pt}improve the offspring $O_1$ and $O_2$ using RLS$^{1,2}$\\
    & \hspace{10pt}for $l$ iterations\\
14  & \hspace{10pt}improve the offspring $O_1$ and $O_2$ using LS$^1$\\
      & \hspace{10pt}until the local optimum is reached\\
15  & \hspace{10pt}if $O_1 \notin P$ then replace the worst individual in $P$ with $O_1$\\
16 & \hspace{10pt}if $O_2 \notin P$ then replace the worst individual in $P$ with $O_2$\\
17 & return the best individual $P_{best}$ in $P$\\
\end{tabular}
\end{center}
\end{table}

\section{Experimental Results}

In this section, we present the experimental of StS AMA and the other algorithms for several real-world network instances. We first describe the experimental protocol for our evaluation. Next, we present the result we obtained for social networks, protein-protein interaction networks and graph studied in network science literature.

\vspace{5pt}
\subsection{Experimental Protocol}

For the experimental evaluation, we have used real-world networks from three different sources. Firstly, we use samples of social networks, including public circles data from \textit{Google+} and social network \textit{Pokec} \cite{TakacPokecAnalysis}. Next, we use protein-protein interaction networks from UCLA database of interacting proteins \cite{SalwinskiDatabaseOfInteractingProteins,XenariosDatabaseOfInteractingProteins2001,XenariosDatabaseOfInteractingProteins,XenariosDatabaseOfInteractingProteins2002}. Last but not least, we perform experiments also for several real-world networks studied in network science literature.

All experiments have been conducted in short-running and long-running forms, with 1 minute and 15 minute time limits. LS$^1$ and ALS$^1$ have been used in a multi-start form, restarting whenever a local optimum has been reached. ARLS$^{1,2}$ has also been used in a multi-start form, with a restart being used after $10^6$ iterations without improvement.

AGA-1PX and AGA-UX have been used with population size $p = 100$ and tournament size $t = 2$. A restart has also been used if $10^6$ individuals without improvement of the best solution found so far have been generated. This restart is also accompanied by adaptation of $p_s$.

StS AMA has also been used with population size $p = 100$ and tournament size $t = 2$. Within StS AMA, RLS$^{1,2}$ has been used with a maximum of $l = 10^6$ iterations.

\vspace{5pt}
\subsection{Results for Social Networks}

\begin{table}
\caption{Comparison of the multi-start variants of the local search algorithms LS$^1$, ALS$^1$, ARLS$^{1,2}$, genetic algorithms AGA-1PX and AGA-UX, and StS AMA for the social networks in short runs with a $1$ minute time limit.}
{\scriptsize
\begin{center}
\begin{tabular}{l l l l l l l}\hline
$G$					& algorithm 		& $\min{\Phi(S)}$		& $E[\Phi(S)]$				& success rate	\\\hline\hline
$soc\_52$				& LS$^1$			& \textbf{0.13108614}			& \textbf{0.13108614}				& 100 / 100	\\
					& ALS$^1$		& \textbf{0.13108614}			& \textbf{0.13108614}				& 100 / 100	\\
					& ARLS$^{1,2}$	& \textbf{0.13108614}			& \textbf{0.13108614}				& 100 / 100	\\
					& AGA-1PX		& \textbf{0.13108614}			& \textbf{0.13108614}				& 100 / 100	\\
					& AGA-UX		& \textbf{0.13108614}			& \textbf{0.13108614}				& 100 / 100	\\
					& StS AMA		& \textbf{0.13108614}			& \textbf{0.13108614}				& 100 / 100	\\
\hline
$gplus\_200$			& LS$^1$			& 0.06158358			& 0.06158358				& 100 / 100	\\
					& ALS$^1$		& \textbf{0.02040816}			& 0.06062967				& 2 / 100		\\
					& ARLS$^{1,2}$	& 0.04854369			& 0.08430937				& 1 / 100		\\
					& AGA-1PX		& 0.06158358			& 0.06411271				& 4 / 100		\\
					& AGA-UX		& 0.06158358			& 0.06453816				& 1 / 100		\\
					& StS AMA		& \textbf{0.02040816}			& \textbf{0.02551749}				& 84 / 100		\\
\hline
$gplus\_500$			& LS$^1$			& 0.03688933			& 0.03688933				& 1 / 100		\\
					& ALS$^1$		& 0.03877551			& 0.04434687				& 1 / 100		\\
					& ARLS$^{1,2}$	& 0.04637097			& 0.07144919				& 1 / 100		\\
					& AGA-1PX		& 0.04608789			& 0.0716302				& 1 / 100		\\
					& AGA-UX		& 0.04772004 			& 0.06875468				& 1 / 100		\\
					& StS AMA		& \textbf{0.02040816}			& \textbf{0.03293719}				& 11 / 100		\\
\hline
$pokec\_500$			& LS$^1$			& 0.02744237			& 0.02952087				& 5 / 100		\\
					& ALS$^1$		& \textbf{0.01345291}			& 0.02891339				& 4 / 100		\\
					& ARLS$^{1,2}$	& 0.03080082			& 0.05740674				& 1 / 100		\\
					& AGA-1PX		& 0.02757916			& 0.06173302				& 1 / 100		\\
					& AGA-UX		& 0.03593556			& 0.06307232				& 1 / 100		\\
					& StS AMA		& \textbf{0.01345291}			& \textbf{0.01345291}				& 100 / 100	\\
\hline
$gplus\_2000$			& LS$^1$			& 0.06439536			& 0.07387353				& 1 / 100		\\
					& ALS$^1$		& 0.06540583			& 0.07352642				& 1 / 100		\\
					& ARLS$^{1,2}$	& 0.0625234			& 0.07842185				& 1 / 100		\\
					& AGA-1PX		& 0.08045977			& 0.1241733				& 1 / 100		\\
					& AGA-UX		& 0.07001123 			& 0.11348164				& 1 / 100		\\
					& StS AMA		& \textbf{0.0494713}			& \textbf{0.05041461}				& 1 / 100		\\
\hline
$pokec\_2000$			& LS$^1$			& 0.0346134			& 0.04484377				& 1 / 100		\\
					& ALS$^1$		& 0.02944942			& 0.04474195				& 1 / 100		\\
					& ARLS$^{1,2}$	& 0.0353544			& 0.06466346				& 1 / 100		\\
					& AGA-1PX		& 0.08342023			& 0.13448438				& 1 / 100		\\
					& AGA-UX		& 0.06587493			& 0.12499968				& 1 / 100		\\
					& StS AMA		& \textbf{0.02360775}			& \textbf{0.02521825}				& 3 / 100		\\
\hline
\end{tabular}
\end{center}
}
\end{table}

In Table 1 and Table 2, the results obtained for social network samples are presented. The first column of these tables identifies the graph, followed by columns denoting the algorithm used, minimum and average conductance $\Phi(S)$ obtained, as well as the success rate, i.e. the number of runs obtaining the best result out of all runs of the algorithm.

We have used the algorithms to solve the problem in a small $52$-vertex social network sample $soc\_52$, samples of public circles data from social network Google+, including $gplus\_200$, $gplus\_500$ and $gplus\_2000$, as well as samples of social network Pokec, including $pokec\_500$, $pokec\_2000$. A larger snapshot of this network is also a part of the SNAP network dataset \cite{LeskovecSnapNetworks}. All of these social network samples are available in an anonymous form\footnote{\url{http://davidchalupa.github.io/research/data/social.html}}.

\begin{table}
\caption{Comparison of the multi-start variants of the local search algorithms ALS$^1$, ARLS$^{1,2}$, and StS AMA for the social networks in longer runs with a $15$ minute time limit.}
{\scriptsize
\begin{center}
\begin{tabular}{l l l l l l l}\hline
$G$					& algorithm 		& $\min{\Phi(S)}$		& $E[\Phi(S)]$				& success rate	\\\hline\hline
$soc\_52$				& ALS$^1$		& \textbf{0.13108614}			& \textbf{0.13108614}				& 30 / 30		\\
					& ARLS$^{1,2}$	& \textbf{0.13108614}			& \textbf{0.13108614}				& 30 / 30		\\
					& StS AMA		& \textbf{0.13108614}			& \textbf{0.13108614}				& 30 / 30		\\
\hline
$gplus\_200$			& ALS$^1$		& \textbf{0.02040816}			& 0.06021106				& 1 / 30		\\
					& ARLS$^{1,2}$	& 0.06233062			& 0.07979315				& 1 / 30		\\
					& StS AMA		& \textbf{0.02040816}			& \textbf{0.02178068}				& 29 / 30		\\
\hline
$gplus\_500$			& ALS$^1$		& 0.03777336			& 0.03972795				& 4 / 30		\\
					& ARLS$^{1,2}$	& 0.04771372			& 0.06992862				& 1 / 30		\\
					& StS AMA		& \textbf{0.034}				& \textbf{0.03401609}				& 29 / 30		\\
\hline
$pokec\_500$			& ALS$^1$		& 0.02744695			& 0.02764317				& 13 / 30		\\
					& ARLS$^{1,2}$	& 0.04347826			& 0.05789981				& 1 / 30		\\
					& StS AMA		& \textbf{0.01345291}			& \textbf{0.01345291}				& 30 / 30		\\
\hline
$gplus\_2000$			& ALS$^1$		& 0.06519208			& 0.0683073				& 1 / 30		\\
					& ARLS$^{1,2}$	& 0.06232454			& 0.07828459				& 1 / 30		\\
					& StS AMA		& \textbf{0.04941531}			& \textbf{0.0499854}				& 6 / 30		\\
\hline
$pokec\_2000$			& ALS$^1$		& 0.03032428			& 0.03623088				& 1 / 30		\\
					& ARLS$^{1,2}$	& 0.03238575			& 0.06166406				& 1 / 30		\\
					& StS AMA		& \textbf{0.02360775}			& \textbf{0.02463325}				& 7 / 30		\\
\hline
$gplus\_10000$		& ALS$^1$		& 0.07737166			& 0.08379242				& 1 / 30		\\
					& ARLS$^{1,2}$	& 0.09005909			& 0.09886034				& 1 / 30		\\
					& StS AMA		& \textbf{0.06645077}			& \textbf{0.0671492}				& 1 / 30		\\
\hline
$pokec\_10000$		& ALS$^1$		& 0.05597898			& 0.08554391				& 2 / 30		\\
					& ARLS$^{1,2}$	& 0.06784962			& 0.08064511				& 1 / 30		\\
					& StS AMA		& \textbf{0.04513233}			& \textbf{0.04656544}				& 1 / 30		\\
\hline
\end{tabular}
\end{center}
}
\end{table}

For $soc\_52$, all algorithms have easily found the same low-conduc\-tance partitioning. However, the algorithms provided much more varied results for larger instances. Interestingly, ALS$^1$ and StS AMA performed significantly better for $gplus\_200$ than the rest of the algorithms. This is most probably due to the adaptive mechanisms of these algorithms. This is also illustrated by Figure 2, showing the difference between the best partitioning found by most of the algorithms for $gplus\_200$ and the partitioning found by ALS$^1$ and StS AMA. This highlights the role of the adaptive approach in the problem, since high-quality solutions to the problem often seem ``asymmetric''. Additionally, the best partitioning found for $gplus\_500$ has the same conductance value as the one found for $gplus\_200$. This indicates that the same community has been identified both in the smaller and the larger sample of the network. However, this solution has been found only by StS AMA for $gplus\_500$.

\newpage
StS AMA has produced the best results also for $gplus\_2000$, \linebreak  $pokec\_500$ and $pokec\_2000$. In addition, one can see that the best conductance has been obtained by StS AMA not only in the best runs, but also on average.

\begin{figure}
\begin{center}
\includegraphics[scale=0.22]{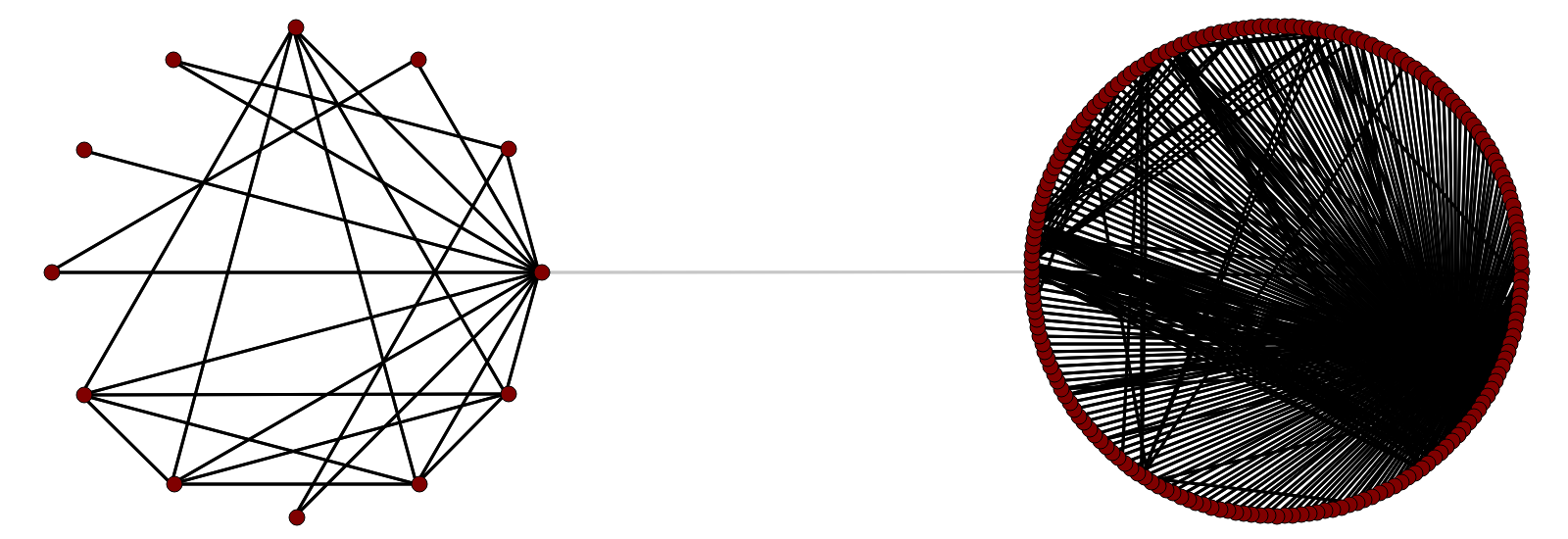}
\includegraphics[scale=0.22]{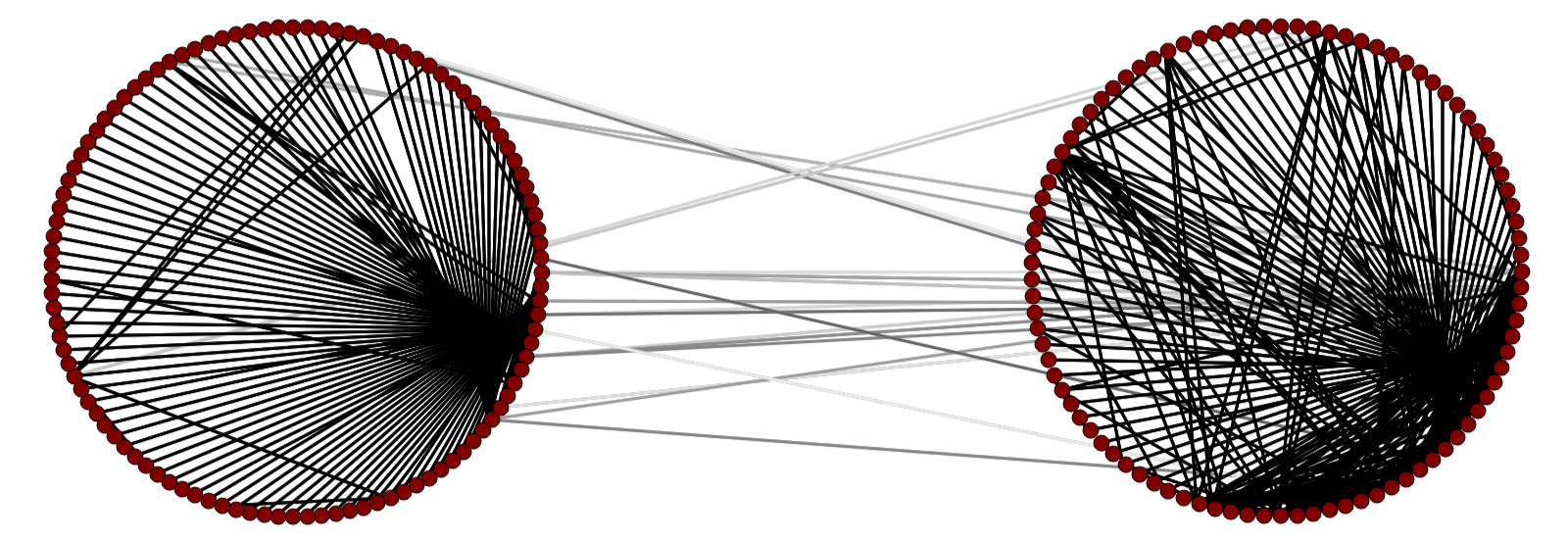}
\caption{Illustrations of two low conductance graph partitionings found for the $200$-vertex sample of publicly available circles data from Google+. The upper partitioning is highly unbalanced, with the conductance value $0.02040816$. The lower partitioning represents a relatively balanced suboptimum with the conductance value $0.06158358$. With the standard way of generating the initial solutions, the algorithms tend to converge to the suboptimum. It is the adaptive initial solution generation, which supports sampling of the unbalanced partitionings.}
\end{center}
\end{figure}

\subsection{Results for Protein-Protein Interaction Networks}

In Table 3 and Table 4, we present the results obtained for protein-protein interaction (PPI) networks. These experiments have been performed for the largest component of each of these networks, since unlike our social network samples, PPI networks do not have to be connected. Graphs with multiple connected components have a trivial solution with zero conductance. 

PPI networks represent the data for the following species. \linebreak $Celeg20160114$ is a PPI network for \textit{Caenorhabditis elegans}, \linebreak $Dmela20160114$ is for the fruit fly, $Ecoli20160114$ is for \textit{Escherichia coli}, \linebreak $Hpylo20160114$ is for \textit{Helicobacter pylori}, $Hsapi20160114$ is a human PPI network, $Mmusc20160114$ is for the house mouse, $Rnorv20160114$ is for the brown rat and $Scere20160114$ is a PPI network for a commonly used species of yeast.

Apart from $Rnorv20160114$, the results of different approaches for PPI networks seem to be much more varied. However, StS AMA has also been identified as by far the most efficient, providing the best results both in the best runs and on average. Intriguingly, the more generous time limit of 15 minutes seemed to give StS AMA improved and somewhat more stable performance. However, there still seems to be large space for performance improvement, either by using large parallel population, or by employing classical graph-theoretical tricks to discover promising regions of the search space.

\subsection{Results for Network Science Graphs}

\begin{table}
\caption{Comparison of the multi-start variants of the local search algorithms LS$^1$, ALS$^1$, ARLS$^{1,2}$, genetic algorithms AGA-1PX and AGA-UX, and StS AMA for the largest connected components of protein-protein interaction networks in short runs with a $1$ minute time limit.}
{\scriptsize
\begin{center}
\begin{tabular}{l l l l l l l}\hline
$G$					& algorithm 		& $\min{\Phi(S)}$		& $E[\Phi(S)]$				& success rate	\\\hline\hline
$Celeg20160114$		& LS$^1$			& 0.1360744			& 0.15048384 				& 1 / 100	\\
					& ALS$^1$		& 0.08558559			& 0.13292784 				& 1 / 100	\\
					& ARLS$^{1,2}$	& 0.13139222			& 0.14561728 				& 1 / 100	\\
					& AGA-1PX		& 0.13117964			& 0.15636037 				& 1 / 100	\\
					& AGA-UX		& 0.1341165			& 0.15054743 				& 1 / 100	\\
					& StS AMA		& \textbf{0.01226994}			& \textbf{0.03331431} 				& 29 / 100	\\
\hline
$Dmela20160114$		& LS$^1$			& 0.30524601			& 0.32650744 				& 1 / 100	\\
					& ALS$^1$		& 0.21733168			& 0.23160832 				& 1 / 100	\\
					& ARLS$^{1,2}$	& 0.23824626			& 0.25221482 				& 2 / 100	\\
					& AGA-1PX		& 0.38336905			& 0.38964162 				& 1 / 100	\\
					& AGA-UX		& 0.32978997			& 0.33968028 				& 1 / 100	\\
					& StS AMA*		& \textbf{0.1559633} 			& \textbf{0.18935739} 				& 1 / 100	\\
\hline
$Ecoli20160114$		& LS$^1$			& 0.44194299			& 0.46075783				& 1 / 100	\\
					& ALS$^1$		& 0.35021218			& 0.38262331 				& 1 / 100	\\
					& ARLS$^{1,2}$	& 0.31840414			& 0.32633622 				& 1 / 100	\\
					& AGA-1PX		& 0.36671548			& 0.40512006 				& 1 / 100	\\
					& AGA-UX		& 0.34065565			& 0.36573979 				& 1 / 100	\\
					& StS AMA		& \textbf{0.06060606}			& \textbf{0.30657497} 				& 1 / 100	\\
\hline
$Hpylo20160114$		& LS$^1$			& 0.16543575			& 0.1878742 				& 1 / 100	\\
					& ALS$^1$		& 0.16543575			& 0.18784113 				& 1 / 100	\\
					& ARLS$^{1,2}$	& 0.17282127			& 0.1934479 				& 1 / 100	\\
					& AGA-1PX		& 0.17429838			& 0.2037738 				& 1 / 100	\\
					& AGA-UX		& 0.17429838			& 0.20141001 				& 1 / 100	\\
					& StS AMA		& \textbf{0.14899926}			& \textbf{0.15361405} 				& 1 / 100	\\
\hline
$Hsapi20160114$		& LS$^1$			& 0.08230694			& 0.08867847 				& 1 / 100	\\
					& ALS$^1$		& 0.06979472			& 0.08627573 				& 1 / 100	\\
					& ARLS$^{1,2}$	& 0.07009483			& 0.08051043				& 1 / 100	\\
					& AGA-1PX		& 0.10188901			& 0.12269489 				& 1 / 100	\\
					& AGA-UX		& 0.08241275			& 0.10327176 				& 1 / 100	\\
					& StS AMA		& \textbf{0.05024438}			& \textbf{0.05558534} 				& 1 / 100	\\
\hline
$Mmusc20160114$		& LS$^1$			& 0.03706222			& 0.04354185 				& 1 / 100	\\
					& ALS$^1$		& 0.03441296			& 0.04331399 				& 1 / 100	\\
					& ARLS$^{1,2}$	& 0.03726083			& 0.04940749 				& 1 / 100	\\
					& AGA-1PX		& 0.04620573			& 0.06044643 				& 1 / 100	\\
					& AGA-UX		& 0.04432505			& 0.0587076 				& 1 / 100	\\
					& StS AMA		& \textbf{0.01428571}			& \textbf{0.02361266} 				& 1 / 100	\\
\hline
$Rnorv20160114$		& LS$^1$			& \textbf{0.00671141}			& \textbf{0.00671141} 				& 100 / 100 \\
					& ALS$^1$		& \textbf{0.00671141}			& \textbf{0.00671141} 				& 100 / 100 \\
					& ARLS$^{1,2}$	& \textbf{0.00671141}			& 0.0120047				& 34 / 100 \\
					& AGA-1PX		& \textbf{0.00671141}			& 0.01036659				& 38 / 100 \\
					& AGA-UX		& \textbf{0.00671141}			& 0.01179875				& 22 / 100 \\
					& StS AMA		& \textbf{0.00671141}			& \textbf{0.00671141} 				& 100 / 100 \\
\hline
$Scere20160114$		& LS$^1$			& 0.45503758			& 0.47404224 				& 1 / 100	\\
					& ALS$^1$		& 0.23874941			& 0.24093083 				& 1 / 100	\\
					& ARLS$^{1,2}$	& 0.23873166			& 0.24034586 				& 1 / 100	\\
					& AGA-1PX		& 0.44679302			& 0.46746535 				& 1 / 100	\\
					& AGA-UX		& 0.34502747			& 0.37700811 				& 1 / 100	\\
					& StS AMA*		& \textbf{0.23821699}			& \textbf{0.23907947} 				& 1 / 100	\\
\hline
\end{tabular}
\end{center}
}
\begin{flushleft}
{\scriptsize
* For these instances, a run of StS AMA took more than 1 minute due to the initial population sampling already taking more than 1 minute.
}
\end{flushleft}
\end{table}

Last but not least, Table 5 and Table 6 present the results obtained for graphs studied in network science literature. These graphs are taken from Newman's network data repository\footnote{\url{http://www-personal.umich.edu/~mejn/netdata/}}. Network $adjnoun$ represents adjective-noun adjacencies in David Copperfield \cite{adjnoun}, $football$	represents matches in a season of American college football league \cite{communitystructure1}, $lesmis$ is a network of character coappearances for Les Miserables \cite{lesmis}, $zachary$ is a network of friendships in a Karate club \cite{zachary}, $celegansneural$ is the neural network for \textit{Ceanorhabditis elegans}	\cite{WattsStrogatzCollectiveDynamicsSmallWorldNetworks}, $dolphins$ is a social network of bottlenose dolphins \cite{LusseauBottlenoseDolphins} and $polbooks$ is a network of political books.

\begin{table}
\caption{Comparison of the multi-start variants of algorithms ALS$^1$, ARLS$^{1,2}$ and StS AMA for the largest connected components of protein-protein interaction networks in long runs with a $15$ minute time limit.}
{\scriptsize
\begin{center}
\begin{tabular}{l l l l l l l}\hline
$G$					& algorithm 		& $\min{\Phi(S)}$		& $E[\Phi(S)]$				& success rate	\\\hline\hline
$Celeg20160114$		& ALS$^1$		& 0.12004018			& 0.13222836 		 		& 1 / 30	\\
					& ARLS$^{1,2}$	& 0.12481645			& 0.14464208 				& 1 / 30	\\
					& StS AMA		& \textbf{0.01226994}			& \textbf{0.01323653}		 		& 23 / 30	\\
\hline
$Dmela20160114$		& ALS$^1$		& 0.21853169			& 0.23147207 	 			& 1 / 30	\\
					& ARLS$^{1,2}$	& 0.2379034			& 0.24385345 				& 1 / 30	\\
					& StS AMA		& \textbf{0.12230216}			& \textbf{0.17148753} 		 		& 1 / 30	\\
\hline
$Ecoli20160114$		& ALS$^1$		& 0.3601381			& 0.38164997 				& 1 / 30	\\
					& ARLS$^{1,2}$	& 0.31797257			& 0.32312285 				& 1 / 30	\\
					& StS AMA		& \textbf{0.05714286}			& \textbf{0.30108943} 				& 1 / 30	\\
\hline
$Hpylo20160114$		& ALS$^1$		& 0.16543575			& 0.17767017 	 			& 2 / 30	\\
					& ARLS$^{1,2}$	& 0.17664449			& 0.19402062 		 		& 1 / 30	\\
					& StS AMA		& \textbf{0.14855876}			& \textbf{0.15223137} 		 		& 2 / 30	\\
\hline
$Hsapi20160114$		& ALS$^1$		& 0.07556946			& 0.08211216 		 		& 1 / 30	\\
					& ARLS$^{1,2}$	& 0.06647116			& 0.07461552 		 		& 2 / 30	\\
					& StS AMA		& \textbf{0.04076645}			& \textbf{0.04409562} 		 		& 1 / 30	\\
\hline
$Mmusc20160114$		& ALS$^1$		& 0.03658946			& 0.03945999 		 		& 1 / 30	\\
					& ARLS$^{1,2}$	& 0.03726083			& 0.04719777 				& 1 / 30	\\
					& StS AMA		& \textbf{0.01242028}			& \textbf{0.01873836} 		 		& 1 / 30	\\
\hline
$Rnorv20160114$		& ALS$^1$		& \textbf{0.00671141}			& \textbf{0.00671141} 				& 30 / 30	\\
					& ARLS$^{1,2}$	& \textbf{0.00671141}			& 0.01732172 		 		& 13 / 30	\\
					& StS AMA		& \textbf{0.00671141}			& \textbf{0.00671141} 				& 30 / 30	\\
\hline
$Scere20160114$		& ALS$^1$		& 0.23933513			& 0.24094871 		 		& 1 / 30	\\
					& ARLS$^{1,2}$	& 0.23846297			& 0.23913658 		 		& 1 / 30	\\
					& StS AMA		& \textbf{0.2376532}			& \textbf{0.23781621} 		 		& 2 / 30	\\
\hline
\end{tabular}
\end{center}
}
\end{table}

For these instances, the algorithms obtained less varied results. No difference in performance of the algorithms was observed for $zachary$ and $polbooks$. In addition, all algorithms obtained the best result also for the other instances. However, StS AMA performed better in terms of its success rate also for these instances.

\begin{table}
\caption{Comparison of the multi-start variants of the local search algorithms LS$^1$, ALS$^1$, ARLS$^{1,2}$, genetic algorithms AGA-1PX and AGA-UX, and StS AMA for the graphs studied in network science in short runs with a $1$ minute time limit.}
{\scriptsize
\begin{center}
\begin{tabular}{l l l l l l l}\hline
$G$					& algorithm 		& $\min{\Phi(S)}$		& $E[\Phi(S)]$				& success rate	\\\hline\hline
$adjnoun$	\cite{adjnoun}			& LS$^1$			& \textbf{0.27830179}			& 0.27867547 				& 78 / 100		\\
					& ALS$^1$		& \textbf{0.27830179}			& 0.27864151				& 80 / 100		\\
					& ARLS$^{1,2}$	& \textbf{0.27830179}			& 0.29172815				& 2 / 100		\\
					& AGA-1PX		& \textbf{0.27830179}			& 0.28204756				& 14 / 100		\\
					& AGA-UX		& \textbf{0.27830179}			& 0.2854969 				& 7 / 100		\\
					& StS AMA		& \textbf{0.27830179}			& \textbf{0.27863774} 				& 82 / 100		\\
\hline
$football$	\cite{communitystructure1}				& LS$^1$			& \textbf{0.10116086}			&\textbf{0.10116086} 				& 100 / 100	\\
					& ALS$^1$		& \textbf{0.10116086}			& \textbf{0.10116086} 				& 100 / 100	\\
					& ARLS$^{1,2}$	& \textbf{0.10116086}			& 0.10267479 				& 95 / 100		\\
					& AGA-1PX		& \textbf{0.10116086}			& 0.11251169 				& 56 / 100		\\
					& AGA-UX		& \textbf{0.10116086}			& 0.11854058 				& 38 / 100		\\
					& StS AMA		& \textbf{0.10116086}			& \textbf{0.10116086} 				& 100 / 100	\\
\hline
$lesmis$	\cite{lesmis}				& LS$^1$			& \textbf{0.12252964}			& \textbf{0.12252964} 				& 100 / 100	\\
					& ALS$^1$		& \textbf{0.12252964}			& \textbf{0.12252964} 				& 100 / 100	\\
					& ARLS$^{1,2}$	& \textbf{0.12252964}			& 0.12256781				& 98 / 100		\\
					& AGA-1PX		& \textbf{0.12252964}			& 0.12303712				& 73 / 100		\\
					& AGA-UX		& \textbf{0.12252964}			& 0.12267752 				& 92 / 100		\\
					& StS AMA		& \textbf{0.12252964}			& \textbf{0.12252964} 				& 100 / 100	\\
\hline
$zachary$	\cite{zachary}				& LS$^1$			&\textbf{0.12820513}			& \textbf{0.12820513} 				& 100 / 100	\\
					& ALS$^1$		& \textbf{0.12820513}			& \textbf{0.12820513} 				& 100 / 100	\\
					& ARLS$^{1,2}$	& \textbf{0.12820513}			& \textbf{0.12820513} 				& 100 / 100	\\
					& AGA-1PX		& \textbf{0.12820513}			& \textbf{0.12820513} 				& 100 / 100	\\
					& AGA-UX		& \textbf{0.12820513}			& \textbf{0.12820513} 				& 100 / 100	\\
					& StS AMA		& \textbf{0.12820513}			& \textbf{0.12820513} 				& 100 / 100	\\
\hline
$celegansneural$	\cite{WattsStrogatzCollectiveDynamicsSmallWorldNetworks}		& LS$^1$			& \textbf{0.17575758}			& \textbf{0.17575758} 				& 100 / 100	\\
					& ALS$^1$		& \textbf{0.17575758}			&\textbf{0.17575758} 				& 100 / 100	\\
					& ARLS$^{1,2}$	& \textbf{0.17575758}			& 0.17894605 				& 37 / 100		\\
					& AGA-1PX		& \textbf{0.17575758}			& 0.18119584 				& 30 / 100		\\
					& AGA-UX		& \textbf{0.17575758}			& 0.18172543				& 41 / 100		\\
					& StS AMA		& \textbf{0.17575758}			& \textbf{0.17575758} 				& 100 / 100	\\
\hline
$dolphins$	\cite{LusseauBottlenoseDolphins}			& LS$^1$			& \textbf{0.06382979}			& \textbf{0.06382979} 				& 100 / 100	\\
					& ALS$^1$		& \textbf{0.06382979}			& \textbf{0.06382979} 				& 100 / 100	\\
					& ARLS$^{1,2}$	& \textbf{0.06382979}			& 0.0761892 				& 87 / 100		\\
					& AGA-1PX		& \textbf{0.06382979}			& \textbf{0.06382979} 				& 100 / 100	\\
					& AGA-UX		& \textbf{0.06382979}			& \textbf{0.06382979} 				& 100 / 100	\\
					& StS AMA		& \textbf{0.06382979}			& \textbf{0.06382979} 				& 100 / 100	\\
\hline
$polbooks$			& LS$^1$			& \textbf{0.04347826}			& \textbf{0.04347826} 				& 100 / 100	\\
					& ALS$^1$		& \textbf{0.04347826}			& \textbf{0.04347826}					& 100 / 100	\\
					& ARLS$^{1,2}$	& \textbf{0.04347826}			& \textbf{0.04347826} 				& 100 / 100	\\
					& AGA-1PX		& \textbf{0.04347826}			& \textbf{0.04347826} 				& 100 / 100	\\
					& AGA-UX		& \textbf{0.04347826}			& \textbf{0.04347826} 				& 100 / 100	\\
					& StS AMA		& \textbf{0.04347826}			& \textbf{0.04347826} 				& 100 / 100	\\
\hline
\end{tabular}
\end{center}
}
\end{table}

\begin{table}
\caption{Comparison of the multi-start variants of algorithms ALS$^1$, ARLS$^{1,2}$ and StS AMA for the graphs studied in network science in long runs with a $15$ minute time limit.}
{\scriptsize
\begin{center}
\begin{tabular}{l l l l l l l}\hline
$G$					& algorithm 		& $\min{\Phi(S)}$		& $E[\Phi(S)]$				& success rate	\\\hline\hline
$adjnoun$	\cite{adjnoun}				& ALS$^1$		& \textbf{0.27830179}			& \textbf{0.27830179}		 		& 30 / 30		\\
					& ARLS$^{1,2}$	& 0.28				& 0.29481774				& 1 / 30		\\
					& StS AMA		& \textbf{0.27830179}			& 0.2784717		 		& 27 / 30		\\
\hline
$football$	\cite{communitystructure1}				& ALS$^1$		& \textbf{0.10116086}			& \textbf{0.10116086} 				& 30 / 30		\\
					& ARLS$^{1,2}$	& \textbf{0.10116086}			& 0.10270889				& 28 / 30		\\
					& StS AMA		& \textbf{0.10116086}			& \textbf{0.10116086} 				& 30 / 30		\\
\hline
$lesmis$	\cite{lesmis}				& ALS$^1$		& \textbf{0.12252964}			& \textbf{0.12252964} 				& 30 / 30		\\
					& ARLS$^{1,2}$	& \textbf{0.12252964}			& \textbf{0.12252964} 				& 30 / 30		\\
					& StS AMA		& \textbf{0.12252964}			& \textbf{0.12252964} 				& 30 / 30		\\
\hline
$zachary$	\cite{zachary}				& ALS$^1$		& \textbf{0.12820513}			& \textbf{0.12820513} 				& 30 / 30		\\
					& ARLS$^{1,2}$	& \textbf{0.12820513}			& \textbf{0.12820513} 				& 30 / 30		\\
					& StS AMA		& \textbf{0.12820513}			& \textbf{0.12820513} 				& 30 / 30		\\
\hline
$celegansneural$	\cite{WattsStrogatzCollectiveDynamicsSmallWorldNetworks}		& ALS$^1$		& \textbf{0.17575758}			& \textbf{0.17575758} 				& 30 / 30		\\
					& ARLS$^{1,2}$	& \textbf{0.17575758}			& 0.18392339		 		& 7 / 30		\\
					& StS AMA		& \textbf{0.17575758}			& 0.17575758 				& 30 / 30		\\
\hline
$dolphins$	\cite{LusseauBottlenoseDolphins}			& ALS$^1$		& \textbf{0.06382979}			& \textbf{0.06382979} 				& 30 / 30		\\
					& ARLS$^{1,2}$	& \textbf{0.06382979}			& 0.07366368 				& 25 / 30		\\
					& StS AMA		& \textbf{0.06382979}			& \textbf{0.06382979} 				& 30 / 30		\\
\hline
$polbooks$			& ALS$^1$		& \textbf{0.04347826}			& \textbf{0.04347826}				& 30 / 30		\\
					& ARLS$^{1,2}$	& \textbf{0.04347826}			& \textbf{0.04347826} 				& 30 / 30		\\
					& StS AMA		& \textbf{0.04347826}			& \textbf{0.04347826} 				& 30 / 30		\\
\hline
\end{tabular}
\end{center}
}
\end{table}

\section{Conclusions and Discussion}

We proposed a \textit{steady-state adaptive memetic algorithm} (StS AMA) for the \textit{minimum conductance graph partitioning problem}. The algorithm combines the steady-state framework with two local search strategies. This includes both randomised local search and systematic local search to ensure that every solution in the population represents a local optimum. Both local search strategies are based on our own efficient \textit{neighbourhood exploration strategy}.

The experimental results were presented for StS AMA, three local search algorithms (systematic local search algorithms LS$^1$, ALS$^1$ and randomised local search algorithm ARLS$^{1,2}$), as well as genetic algorithms with one-point and uniform crossovers (AGA-1PX and AGA-UX). The experiments were performed on real-world networks, including social network samples, protein-protein interaction networks and graphs studied in network science literature.

These results identified StS AMA as the most robust strategy to solve the minimum conductance problem. We have also identified that the performance gap between StS AMA and the other algorithms seems to become wider as the instances get larger. The largest gaps have been identified for large social network samples with $2000$ vertices, as well as for large protein-protein interaction networks.

However, several problems remain open. It is not yet clear whe\-ther improving moves can be identified in $\mathcal{O}(1)$ time for the minimum conductance problem, similarly to max-SAT or NK-landscapes \cite{ConstantTimeSteepestDescent,BestImprovingVersusFirstImproving}. Performance of StS AMA for large problem instances also suggests that improved results may be obtained by a parallel or a distributed variant of the algorithm. Last but not last, a hybrid approach, combining StS AMA with classical graph-theoretical techniques can also be a way to improve the performance of the algorithm.

\bibliography{common}{}
\bibliographystyle{plain}

\end{document}